\documentclass[aps,showpacs,nofootinbib]{revtex4}
\usepackage{epsfig}
\usepackage{latexsym}
\usepackage{graphics}
\usepackage{graphicx,pstricks}
\usepackage{amsmath,amstext,amsthm,amssymb}
\usepackage{wrapfig}

\usepackage{ulem}

\begin{document}

\title{Influence of correlations between yields on the chemical freeze-out temperature}
\author{Dariusz Prorok}
\email{dariusz.prorok@ift.uni.wroc.pl}
\affiliation{Institute of Theoretical Physics, University of
Wroc{\l}aw,\\ Pl.Maksa Borna 9, 50-204  Wroc{\l}aw, Poland}
\date{April 28, 2021}

\begin{abstract}
 A statistical (thermal) model is applied to the description of hadron yields measured at central nucleus-nucleus collisions at the top RHIC energy $\sqrt{s_{NN}} = 200$ GeV and the LHC energy $\sqrt{s_{NN}} = 2.76$ TeV. In contrast to the previous analyzes a more general form of the least squares test statistic is used, which takes into account also possible correlations between different species of yields.  In addition to hadrons also light nuclei are included in  the  fits. Because of the lack of  data, a toy model is constructed where the correlation coefficients are just free parameters.  Because of this  the presented considerations are speculative and should be treated as an imperfect illustration of the problem. Within these limitations it is impossible to formulate any definite conclusion, it might be only mentioned that for the considered examples the dependence of the freeze-out temperature and baryon chemical potential on correlations turned out to be weak.
\end{abstract}

\maketitle


\section{Introduction}
\label{intro}

The Quantum Chromodynamics (QCD) -- the  established  theory of strong interactions, predicts a phase transition from a system of hadrons to  the  system of quarks and gluons. This can happen at extremely high temperatures  and/or  densities of the system. The circumstances necessary for the creation of such a phase  (called the Quark Gluon Plasma) can be established in the laboratory presently (for a review of the subject, see Ref.~\cite{Leupold:2011zz}) e.g.\ in high energy collisions of heavy ions. Matter created during such a collision, extremely dense and hot, is
initially compressed in the volume of a narrow disk of the size of the nucleus
diameter. Then, because of the huge pressure, it expands rapidly and cools down
simultaneously. Its evolution  is described usually with the help of
relativistic hydrodynamics \cite{Huovinen:2013wma}. During this rapid expansion matter
undergoes a phase transition into the hadron gas phase. We assume that the hadron gas continues to
evolve hydrodynamically, for some time after the moment of transition. The expansion is causing gas to become
increasingly thinner, so when the average free path of its components becomes
comparable with the size of the system, one can no longer treat it as a collective system. After this moment, referred to as \textit{freeze-out}, the gas breaks down into
freely escaping particles, which can be detected. Basically, two moments during
freezing can be distinguished: \textit{chemical freeze-out} when all inelastic
interactions disappear and, at lower temperature, \textit {kinetic freeze-out},
when also elastic interactions disappear. The measured hadron yields reflect the
corresponding hadron abundances present during the chemical freeze-out. Multiplicities can be described in the framework of a grand-canonical ensemble (GCE) by only three independent parameters: the chemical freeze-out temperature $T_{ch}$, the baryochemical potential $\mu_B$, and the system volume at the moment of
freeze-out $V$ \cite{BraunMunzinger:2003zd,Floris:2014pta}. This idea is
the basis of the application of the Statistical Model (SM) to the production of particles in
heavy-ion collisions. It has been proposed, that also light nuclei can be incorporated into this scheme \cite{Siemens:1979dz,BraunMunzinger:1994iq,BraunMunzinger:2001mh}. This is because the entropy per baryon, which is set at the chemical freeze-out, is crucial in this case, not the fact that the nuclei binding energies are much smaller than the temperature \cite{Siemens:1979dz}. Therefore, the measured light nucleus yields (or their ratios) are also included into fits in the present analysis.

\section{The model}
\label{modeldescr}

Let $\vec{Y}$ be an $N$-dimensional Gaussian random variable of measured yields with known covariance matrix $\textsc{C} = [C_{ij}]$ ($\textsc{C}$ must be positive definite) but unknown expectation values. The $\vec{Y}^{th}$ represents theoretical predictions for yields, i.e. $Y_i^{th} = V \cdot n_i(T_{ch},\mu_B)$, where $V$ is the volume of the system and $n_i(T_{ch},\mu_B)$ is the thermal density with contributions from resonance decays,

\begin{equation}
n_i(T_{ch}, \mu_{B}) = n_{i}^{prim}(T_{ch}, \mu_{B}) + \sum_{a}
\varrho(i,a)\; n_{a}^{prim}(T_{ch}, \mu_{B}) \;, \label{nchj}
\end{equation}
where $n_{i}^{prim}(T_{ch}, \mu_{B})$ is the thermal density of particle
species $i$ at the freeze-out, $\varrho(i,a)$ is the final fraction of particle species $i$ which can
be received from all possible decays (cascades) of particle $a$, and the sum is over all kinds of resonances in the hadron gas. Here, the system at the freeze-out is a point-like hadron gas, which consists of all stable hadrons and light-flavour resonances up to $\sim$2 GeV \cite{Agashe:2014kda}. Then one defines the least-squares (LS) statistic as \cite{Cowan:1998ji,Barlow:1989}:

\begin{equation}
\chi_{LS}^{2}(\vec{Y};V,T_{ch},\mu_B) = \sum_{i,j=1}^{N} (Y_i-Y_i^{th}
) [\textsc{C}^{-1}]_{ij} (Y_j - Y_j^{th}) \;. \label{ChiLSdef}
\end{equation}
However, if one treats yields as independent Gaussian random variables with known variances $\sigma_i^2$, then the LS statistic takes the form:

\begin{equation}
\chi_{LS}^{2}(\vec{Y};V,T_{ch},\mu_B) = \sum_{i=1}^{N} \frac{(Y_i-Y_i^{th})^2}{\sigma_i^2}  \;. \label{ChiLSdef2}
\end{equation}
Both test statistics, Eq.~(\ref{ChiLSdef}) and Eq.~(\ref{ChiLSdef2}), can be obtained from the principle of maximum likelihood \cite{Cowan:1998ji,Barlow:1989}. In the case of Eq.~(\ref{ChiLSdef}), the likelihood is given by $N$-dimensional Gaussian probability distribution function (p.d.f.) taken at $\vec{Y}$ -- a single measurement of the $N$-dimensional random Gaussian variable. Thus the significance of Eq.~(\ref{ChiLSdef}) is exactly the same
as Eq.~(\ref{ChiLSdef2}). To determine the optimal values of parameters one has to minimize the function given by Eq.~(\ref{ChiLSdef}) or Eq.~(\ref{ChiLSdef2}) with respect to $V,\;T_{ch}$ and $\mu_B$. Moreover, one should remember that Eq.~(\ref{ChiLSdef}) (or Eq.~(\ref{ChiLSdef2}))
defines not a normal function but a test statistics that is the
function which is a random variable. As a random variable this
function has its own distribution, usually unknown. Not going
into details, to make a decision one has to assess the
goodness-of-fit, e.g. calculating the $\emph{p-value}$ of the fit. Without knowledge of the distribution of the test statistic it
is impossible, however, there is a mathematical theorem which
says when this distribution becomes the true $\chi^2$
distribution. Namely, if

\begin{enumerate}

\item $(Y_1,Y_2,...,Y_N)$ is an $N$-dimensional Gaussian
random variable with known covariance matrix $\textsc{C}$ or
$(Y_1,Y_2,...,Y_N)$ are independent Gaussian random variables
with known variances $\sigma_i^2$;

\item the hypothesis
$Y_i^{th}(\theta_1,...,\theta_m)$ is linear in the parameters
$\theta_i$; and

\item the hypothesis is correct,
\end{enumerate}
then the test statistic $\chi_{LS,min}^{2}$ is distributed according to a
$\chi^{2}$ distribution with $n_d = N-m$ degrees of freedom.

If the hypothesis $Y_i^{th}(\theta_1,...,\theta_m)$ is
nonlinear in the parameters, the exact distribution of
$\chi_{LS,min}^{2}$ is not known. However, asymptotically (when
$N \longrightarrow \infty$) the distribution of
$\chi_{LS,min}^{2}$ approaches a $\chi^{2}$ distribution as
well (\cite{Frodesen:1979fy}, p. 287; \cite{Roe:1992zz}, p. 147) Thus when at least assumptions 1 and 3
are fulfilled and the sample size is large one can consider the
$\chi_{LS,min}^{2}$ test statistic to be $\chi^{2}$ distributed.
The expectation value of a random variable $Z$ distributed
according to the $\chi^{2}$ distribution with $n_d$ degrees of
freedom is $E[Z] = n_d$ and the variance $V[Z] = 2n_d$. As a
result one expects to obtain
$\chi_{LS,min}^{2} \approx n_d$ in a "reasonable" experiment. Therefore for the test
statistic $t_{\chi^{2}} = \chi_{LS,min}^{2}$ the decision
boundary $t_{\chi^{2},cut} = E[\chi_{LS,min}^{2}] = n_d$ is
chosen. Usually the so-called ``reduced $\chi^{2}$'' is reported,
defined as $\chi_{LS,min}^{2}/n_d$, so for
$\chi_{LS,min}^{2}/n_d$ the decision boundary is just one. It
must be stressed here that this choice is the consequence of
the fact that the $\chi_{LS,min}^{2}$ test statistic is
$\chi^{2}$ distributed. If the distribution of the
$\chi_{LS,min}^{2}$ is not known at all (e.g. one of the
assumptions 1 or 3 is not fulfilled or the sample size is
small), this choice is arbitrary -- usually based on common belief
rather than any justification. So the main message from all
of this, is that to make any reasonable assessment of the
optimal values of parameters, one must know the elements of the
covariance matrix in advance and these elements must be
constant numbers. In particular they can not depend on the
parameters. Moreover, off-diagonal elements of the covariance matrix
describe all the possible correlations or more precisely -- if
one applies all of this to physics, they represent the result
of the sum over all possible sources of correlations. The same concerns diagonal elements of the
covariance matrix --  they are variances so, in fact, they represent the
true fluctuations. In practice they are often replaced
by squares of experimental errors. In this way the
contribution from physical fluctuations is neglected in the standard so-called "chi-squared" minimization. The same  applies to  off-diagonal elements of the covariance matrix, so in general
any attempt to include somehow physically justified correlations
and fluctuations, i.e. correlations and fluctuations which are
estimated within a physical model, breaks the method, because
usually such estimates depend on parameters of the model. Thus,
the off-diagonal elements of the covariance matrix should be
also estimated by experimentalists.

All fits to yields done so far, have been performed with the help of the LS statistic given by Eq.~(\ref{ChiLSdef2}), but the obvious pitfall of this simplification is that the possible experimental correlations between yields are neglected. This could be justified in the case of identified hadrons (pions, kaons and (anti-)protons) but not in the case of resonances. The latter are not measured directly, but via they decay products -- the above-mentioned identified hadrons. That is $\phi \rightarrow K^- + K^+$, $K^0_S \rightarrow \pi^+ + \pi^-$, $\Lambda \rightarrow p + \pi^-$, $\Xi^- \rightarrow \Lambda + \pi^-$ and $\Omega^- \rightarrow \Lambda + K^-$ with subsequent decay $\Lambda \rightarrow p + \pi^-$. It means that they are reconstructed from pions, kaons and (anti-)protons, which have been extracted with the help of some techniques from the whole samples of these particles. Therefore, the resonances have to be correlated with their daughter particles. The dominant source of these correlations is purely statistical -- they appear always when two measurements use the same subset of data, i.e. the set of data used to construct the first measurement is not disjoint with the set of data used to construct a second measurement (see section 7.6.2 in \cite{Cowan:1998ji}). Let's consider as an example the instructive case of $\phi$ meson. The meson is reconstructed via its decay channel $\phi \rightarrow K^- + K^+$. The first step is to make all possible unlike-charge pairs of kaons from the same event. But $K^-$'s and $K^+$'s which contribute to these pairs belong also to whole samples of $K^-$ and $K^+$, respectively. These whole samples are used to obtain corresponding yields, so there are subsets of $K^-$ and $K^+$ which are used simultaneously to obtain yields of $\phi$ meson and kaons. Certainly, this happens in the case of ALICE hadron data \cite{Abelev:2013vea,Abelev:2013xaa,ABELEV:2013zaa,Abelev:2014uua} because they all were recorded during the first Pb-Pb run at the LHC in the Autumn of 2010. In the RHIC case situation is different because the hadron data analysed here were collected during two runs, in 2001 \cite{Abelev:2008ab,Adams:2006ke} and 2004 \cite{Abelev:2008aa,Agakishiev:2011ar}. However, it is not clear whether the later data include the earlier samples or not. Anyway, to do a consistent analysis of both collider cases the present approach is applied also to the RHIC case.

Another possible contribution to the correlations between yields could be of  physical origin as described within the GCE in \cite{Torrieri:2005va,Torrieri:2006xi,Torrieri:2007ca}. However, the main pitfall of attributing the correlations entirely to these physical effects is that a centrality class is not an ensemble. The ensemble consists of a large number of systems prepared exactly in the same way, which is not true for events belonging to the same centrality class. Note that because of this reason, comparison of experimental yields with those calculated within the GCE is also an approximation.

Since the corresponding elements of the covariance matrix are not given in general (precisely, they should be somehow estimated by experimentalists), they have to be modeled. It is assumed here, that the only non-zero off-diagonal elements of the covariance matrix are those which are between resonances and their final daughter pions, kaons or (anti-)protons and the resonances and subsequent resonances in the case of a cascade. These are: $(\phi, K^-)$, $(\phi,K^+)$, $(K^0_S,\pi^+)$, $(K^0_S,\pi^-)$, $(\Lambda,p)$, $(\Lambda,\pi^-)$,
$(\Xi^-,\Lambda)$, $(\Xi^-,\pi^-)$, $(\Omega^-,\Lambda)$, $(\Omega^-,K^-)$, $(\Omega^-,\pi^-)$, $(\Xi^-,p)$, $(\Omega^-,p)$ and corresponding antiparticle pairs. Because (anti-)hypertriton $(^3_{\bar{\Lambda}}\bar{\textrm{H}})$ $^3_{\Lambda}\textrm{H}$ is measured via its mesonic decay $(^3_{\bar{\Lambda}}\bar{\textrm{H}} \rightarrow ^3\!\!\bar{\textrm{H}} + \pi^+)$ $^3_{\Lambda}\textrm{H} \rightarrow ^3\!\!\textrm{H} + \pi^-$ also correlations of pairs $(^3_{\Lambda}\textrm{H},^3\!\!\textrm{H})$, $(^3_{\Lambda}\textrm{H},\pi^-)$ and corresponding antipairs are assumed to be non-zero. One must also take into account, that in some cases contributions to a final yield from weak decays are subtracted, which certainly diminishes corresponding correlations. In the case of Pb-Pb collisions at $\sqrt{s_{NN}} = 2.76$ TeV (ALICE), the contribution from the weak decays concerns mainly (anti-)protons  \cite{Abelev:2013vea,Milano:2012eea}, hence secondary (anti-)protons from primordial and decay $\Lambda$($\bar{\Lambda}$)'s are subtracted. In the case of Au-Au collisions at $\sqrt{s_{NN}} = 200$ GeV (STAR), pions from decays of $K^0_S$ and $\Lambda$ are subtracted \cite{Abelev:2008ab}. In this case of the top RHIC energy only data reported by the STAR Collaboration will be analyzed because mixing data from different detectors are biased by an additional systematic error, which is hard to estimate.

From the definition of the correlation coefficient $\rho_{ij}$ one has

\begin{equation}
C_{ij} = \rho_{ij} \sigma_i \sigma_j
\;, \label{Covariance}
\end{equation}
where $\sigma_i$ and  $\sigma_j$ are standard deviations of $Y_i$ and $Y_j$, respectively. Because off-diagonal elements of the covariance matrix are not known, it is assumed here, for simplicity, that, in fact, there are only 2 different correlation coefficients: one connected with the species which are corrected for weak decays and the other for all the rest. So, for the ALICE case $\rho_1$ is defined as the correlation coefficient for all cases where a proton is a daughter particle and $\rho_2$ for others. For the STAR case, $\rho_1$ is defined as the correlation coefficient for all cases where a pion is a daughter particle and $\rho_2$ for others. As far as $\sigma_i$'s are concerned, they are replaced by experimental errors (statistical and systematic) in the following. The $\rho_1$ and $\rho_2$ are free parameters and are not fitted here (in fact they can not be fitted, because the covariance matrix has to be known by definition of the LS method \cite{Cowan:1998ji,Barlow:1989}). However, it has turned out that they are bounded from above by values significantly less than 1 (these maximal values are given in Tables~\ref{Table1}-\ref{Table2} \footnote{They were determined in the following way: first, values $\rho_1 = \rho_2 = 1$ were substituted into the covariance matrix, then, because the matrix turned out not to be positive definite, $\rho_1$ and $\rho_2$ were diminished gradually keeping them equal, until the matrix become positive definite. Second, $\rho_2$ was increased keeping $\rho_1$ constant until the matrix ceases to be positive definite. Then $\rho_1$ was increased with $\rho_2$ fixed at the value determined in the previous step. If $\emph{p-value}$ was smaller than 1 \%, then the whole procedure was repeated but with the criterion that $\emph{p-value}$ must be at least 1 \%.}). This is because for greater values either the covariance matrix is no longer positive definite and the minimization of the LS statistic, Eq.~(\ref{ChiLSdef}), doesn't makes sense (the STAR case) or $\emph{p-value}$ becomes smaller than 1 \% - the significance level assumed here (the ALICE case). In the first case, it does not mean that correlations can not be greater. They can, but then the yields do not compose a multivariate Gaussian random variable. One should also notice that maximal values of $\rho_1$ are much smaller than those of $\rho_2$. This is reasonable, because $\rho_1$ represents cases where products of weak decays are subtracted, what should diminish correlations. This is because not full corrections for weak decays are performed by experimentalists. In considered cases only protons from $\Lambda$ (ALICE) and pions from decays of $K^0_S$ and
$\Lambda$ (STAR) are subtracted. Moreover, the amount of subtracted
particles is an estimate obtained with the help of Monte Carlo
simulations, and such subtraction is not perfect.

\begin{figure}
\includegraphics[width=0.85\textwidth]{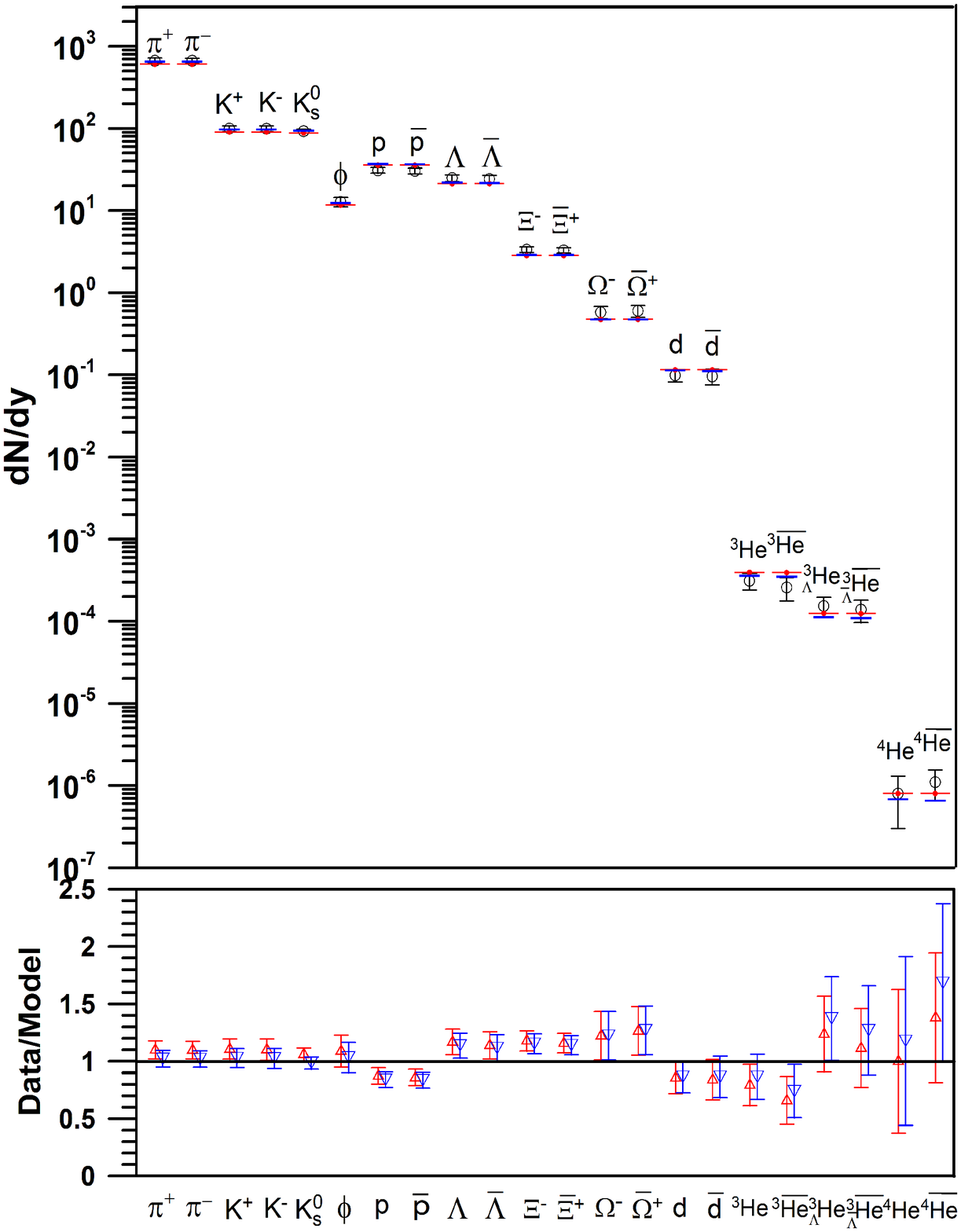}
\caption{\label{fig.1} Yields of hadrons and light nuclei measured in Pb-Pb collisions at $\sqrt{s_{NN}} = 2.76$ TeV for 0-10\% centrality class (open circles) compared to the predictions of the statistical hadronization model with correlations included (red bars and triangles) and without correlations (blue bars and upside down triangles), errors are sums of statistical and systematic components added in quadrature. Data are from \protect\cite{Abelev:2013vea,Abelev:2013xaa,ABELEV:2013zaa,Abelev:2014uua,Adam:2015yta,Adam:2015vda,Acharya:2017bso}. }
\end{figure}

\begin{figure}
\includegraphics[width=0.85\textwidth]{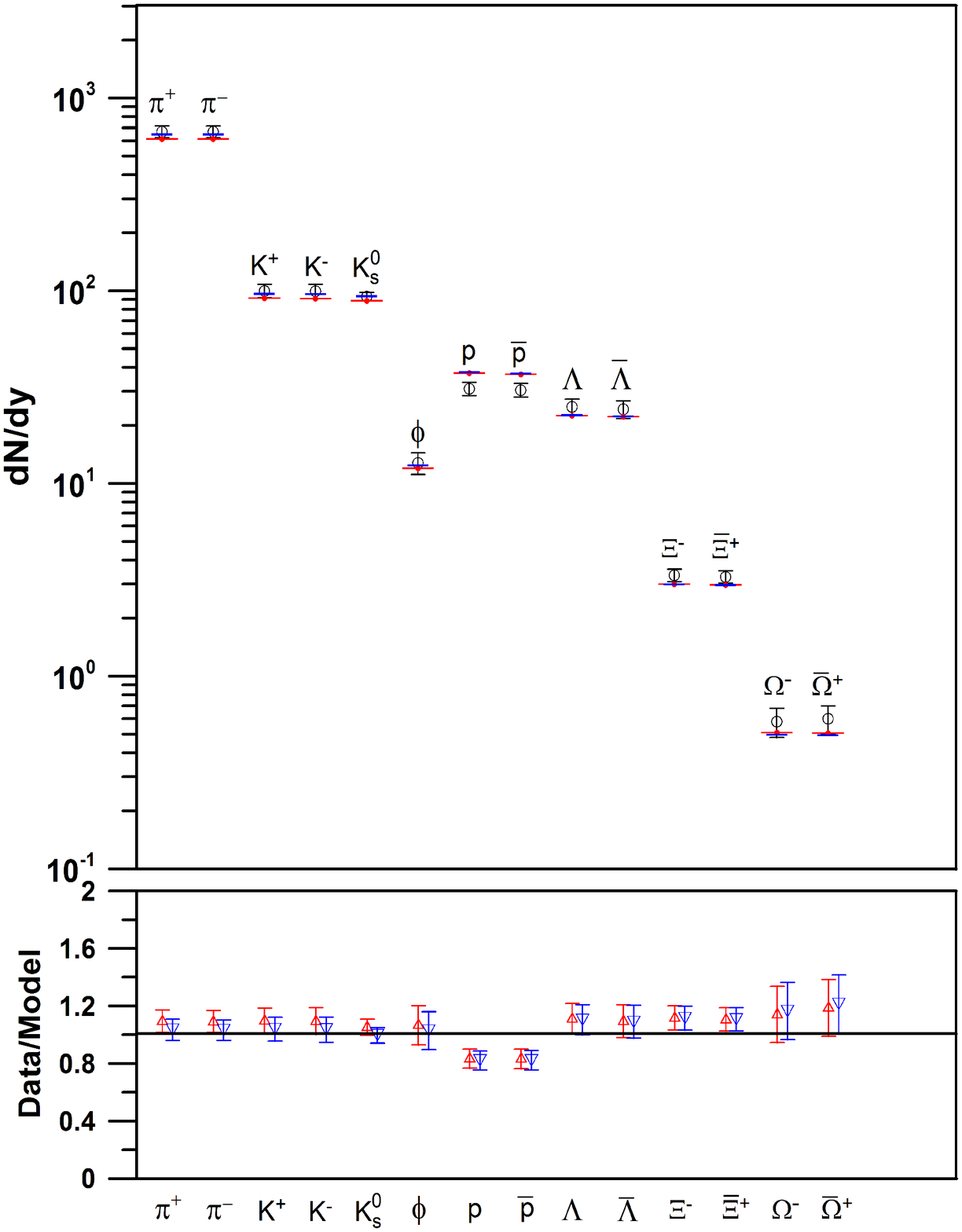}
\caption{\label{fig.2} Yields of hadrons measured in Pb-Pb collisions at $\sqrt{s_{NN}} = 2.76$ TeV for 0-10\% centrality class (open circles) compared to the predictions of the statistical hadronization model (light nuclei excluded in fits) with correlations (red bars and triangles) and without correlations (blue bars and upside down triangles), errors are sums of statistical and systematic components added in quadrature. Data are from \protect\cite{Abelev:2013vea,Abelev:2013xaa,ABELEV:2013zaa,Abelev:2014uua}. }
\end{figure}

\section{Results}
\label{result}

\begin{figure}
\includegraphics[width=0.85\textwidth]{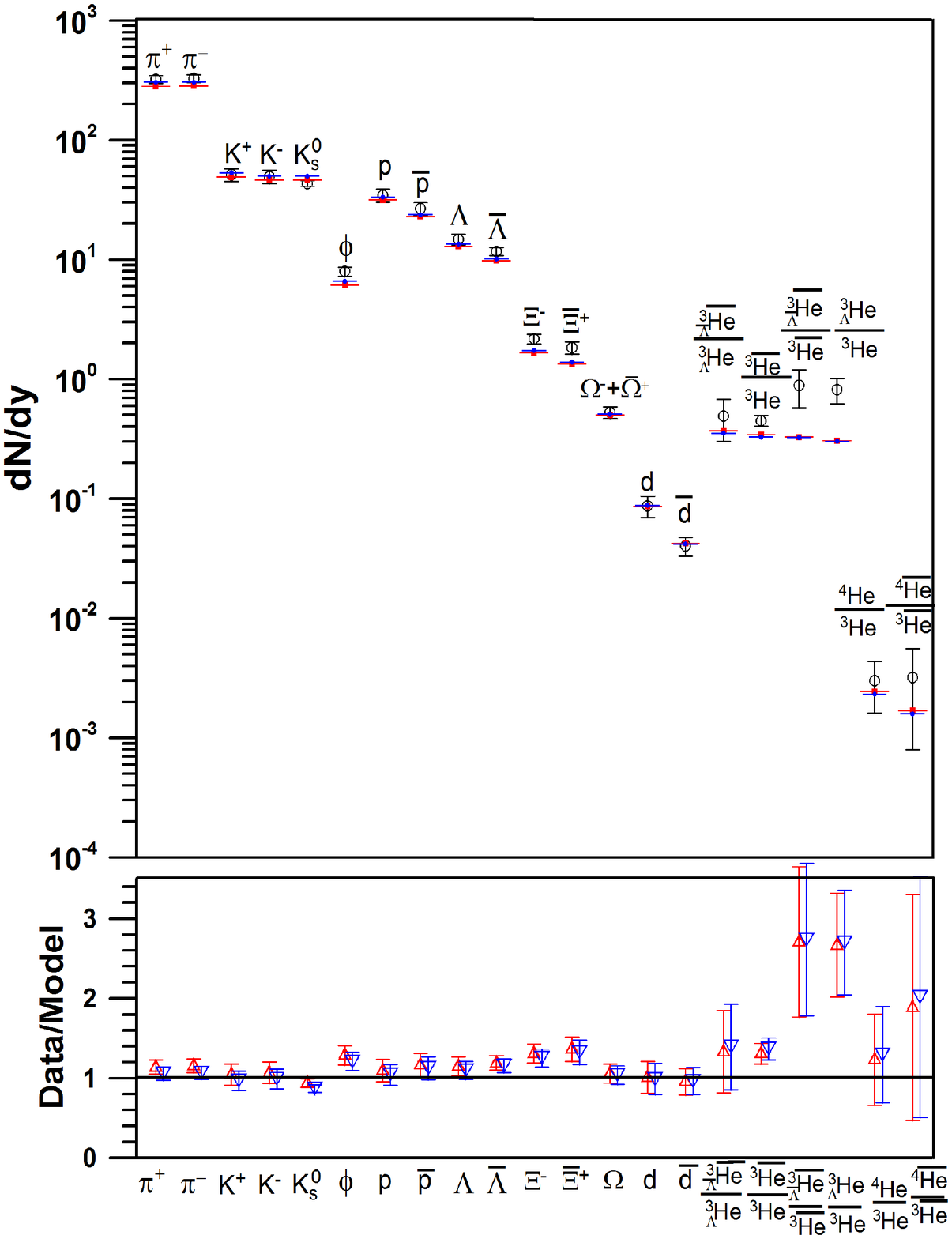}
\caption{\label{fig.3} Yields of hadrons and yield ratios of light nuclei measured in Au-Au collisions at $\sqrt{s_{NN}} = 200$ GeV for 0-5\% centrality class (open circles) compared to the predictions of the statistical hadronization model with correlations included (red bars and triangles) and without correlations (blue bars and upside down triangles), errors are sums of statistical and systematic components added in quadrature. Data are from \protect\cite{Abelev:2008ab,Adams:2006ke,Abelev:2008aa,Agakishiev:2011ar,Agakishiev:2011ib,Abelev:2010rv,Adam:2019wnb}.}
\end{figure}

\begin{figure}
\includegraphics[width=0.85\textwidth]{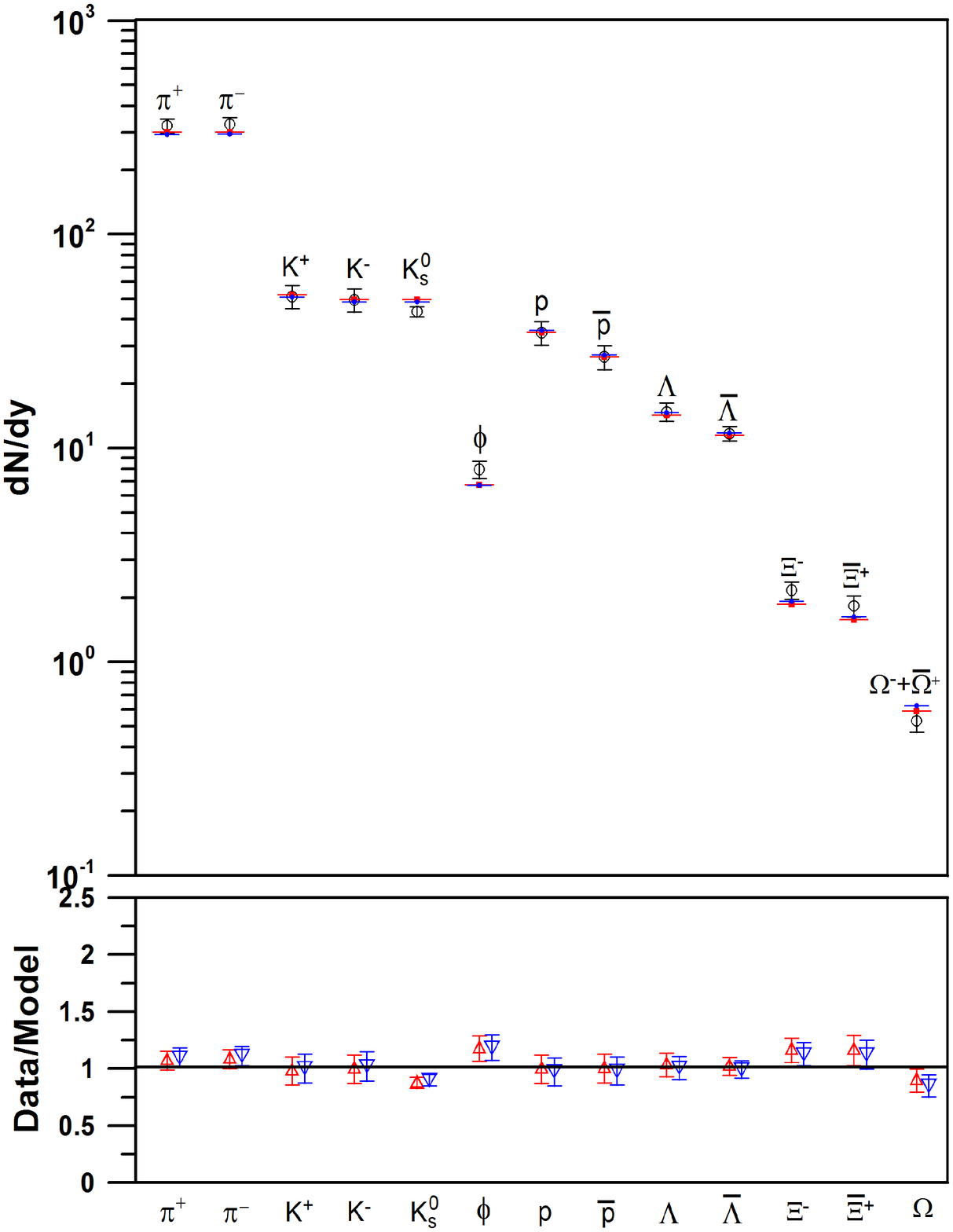}
\caption{\label{fig.4} Yields of hadrons measured in Au-Au collisions at $\sqrt{s_{NN}} = 200$ GeV for 0-5\% centrality class (open circles) compared to the predictions of the statistical hadronization model with correlations included (red bars and triangles) and without correlations (blue bars and upside down triangles), errors are sums of statistical and systematic components added in quadrature. Data are from \protect\cite{Abelev:2008ab,Adams:2006ke,Abelev:2008aa,Agakishiev:2011ar}.}
\end{figure}

The results of fits are presented in Tables~\ref{Table1} and \ref{Table2} and depicted in Figs.\,\ref{fig.1}-\ref{fig.4}. The dependence of the freeze-out temperature on the correlation coefficients has turned out to be weak (differences in temperature in the considered ranges of correlation coefficients are 2\% at most), therefore only cases with maximal possible values of these coefficients are reported. The interesting observation is that, when correlations are taken into account, the chemical freeze-out temperatures determined in the RHIC-STAR case and the LHC-ALICE case agree within errors and are 158 MeV with light nuclei included and about 160 MeV without. Note that for zero correlations the temperature obtained for the LHC-ALICE, $T_{ch} = 155.8 \pm 1.2$ MeV, agrees well with the corresponding temperature, $T_{ch} = 156.5 \pm 1.5$ MeV, reported in \cite{Andronic:2017pug} (the fit with $\chi^{2}/n_{dof} = 1.62$ and $\emph{p-value} = 4.7\;\%$). Also the baryochemical potential agrees, here $\mu_B = 0.8 \pm 3.7$ MeV, whereas $\mu_B = 0.7 \pm 3.8$ MeV in \cite{Andronic:2017pug}. Only the volume disagrees, here $V = 4198 \pm 307$ fm$^3$ ($V = 5280 \pm 410$ fm$^3$ in \cite{Andronic:2017pug}), but this can be explained by the fact that point-like particles are assumed in present model whereas a hard-sphere excluded volume approach is applied in \cite{Andronic:2017pug}.

The overall agreement with the data is reasonable and the quality of fits is acceptable, however with increasing correlations the quality worsens. As it was already pointed out in the literature \cite{Floris:2014pta,Andronic:2017pug,Stachel:2013zma} the biggest discrepancy concerns (anti-)protons in the LHC-ALICE case, here a deviation of 2$\sigma$ is obtained for non-zero correlations (it is significantly lower than 2.7$\sigma$ reported in \cite{Andronic:2017pug}) and 2.4$\sigma$ for 0 correlations. For the RHIC-STAR case the biggest discrepancy of the model predictions with the data is seen for the ratios of the (anti-)hypertriton to (anti-)helium-3, where a deviation of (1.8) 2.5$\sigma$ is obtained.

\begin{table*}
\caption{\label{Table1} Fit results for Pb-Pb collisions at $\sqrt{s_{NN}} = 2.76$ TeV and the measurement at central rapidity, $\mid y \mid < 0.5$. }
\begin{ruledtabular}
\begin{tabular}{ccccccc} \hline
 \multicolumn{7}{c}{\textbf{with light nuclei}, $n_{dof}=19$}
\\
\hline\hline
$\rho_1$ & $\rho_2$ & $T_{ch}$ (MeV) & $\mu_B$ (MeV) & $V$ (fm$^3$)& $\chi^{2}/n_{dof}$ & $\emph{p-value}\;(\%)$
\\
\cline{1-7}
 0.140 & 0.341 & $157.9 \pm 1.1$ & $0.08 \pm 3.85$ & $3544.3 \pm 232.5$ & 1.90 & 1.0
\\
\hline\hline
 \multicolumn{7}{c}{parameter correlations}
\\
\cline{1-7}
 \multicolumn{3}{c}{($T_{ch},\mu_B$)} & \multicolumn{2}{c}{($T_{ch},V$)} & \multicolumn{2}{c}{($\mu_B,V$)}
\\
\cline{1-7}
 \multicolumn{3}{c}{-0.052} & \multicolumn{2}{c}{-0.912} & \multicolumn{2}{c}{0.033}
\\
\hline\hline
  $\rho_1$ & $\rho_2$ & $T_{ch}$ (MeV) & $\mu_B$ (MeV) & $V$ (fm$^3$)& $\chi^{2}/n_{dof}$ & $\emph{p-value}\;(\%)$
\\
\hline
0.0 & 0.0 & $155.8 \pm 1.2$ &  $0.79 \pm 3.65$ & $4198.3 \pm 307.2$ & 1.52 & 6.8
\\
\hline\hline
 \multicolumn{7}{c}{parameter correlations}
\\
\hline
 \multicolumn{3}{c}{($T_{ch},\mu_B$)} & \multicolumn{2}{c}{($T_{ch},V$)} & \multicolumn{2}{c}{($\mu_B,V$)}
\\
\hline
 \multicolumn{3}{c}{-0.078} & \multicolumn{2}{c}{-0.955} & \multicolumn{2}{c}{0.064}
\\
\hline\hline
 \multicolumn{7}{c}{\textbf{without light nuclei}, $n_{dof}=11$}
\\
\hline\hline
  $\rho_1$ & $\rho_2$ & $T_{ch}$ (MeV) & $\mu_B$ (MeV) & $V$ (fm$^3$)& $\chi^{2}/n_{dof}$ & $\emph{p-value}\;(\%)$
\\
\cline{1-7}
 0.171 & 0.370 & $159.9 \pm 1.6$ & $1.13 \pm 5.93$ & $3264.7 \pm 286.0$ & 2.25 & 1.0
\\
\hline\hline
 \multicolumn{7}{c}{parameter correlations}
\\
\hline
 \multicolumn{3}{c}{($T_{ch},\mu_B$)} & \multicolumn{2}{c}{($T_{ch},V$)} & \multicolumn{2}{c}{($\mu_B,V$)}
\\
\hline
 \multicolumn{3}{c}{0.010} & \multicolumn{2}{c}{-0.948} & \multicolumn{2}{c}{-0.011}
\\
\hline\hline
 $\rho_1$ & $\rho_2$ & $T_{ch}$ (MeV) & $\mu_B$ (MeV) & $V$ (fm$^3$)& $\chi^{2}/n_{dof}$ & $\emph{p-value}\;(\%)$
\\
\hline
 0.0 & 0.0 & $157.4 \pm 2.0$ & $1.39 \pm 5.08$ & $3858.8 \pm 425.0$ & 2.02 & 2.3
\\
\hline\hline
 \multicolumn{7}{c}{parameter correlations}
\\
\hline
 \multicolumn{3}{c}{($T_{ch},\mu_B$)} & \multicolumn{2}{c}{($T_{ch},V$)} & \multicolumn{2}{c}{($\mu_B,V$)}
\\
\hline
 \multicolumn{3}{c}{-0.003} & \multicolumn{2}{c}{-0.979} & \multicolumn{2}{c}{0.001}
\\
\hline
\end{tabular}
\end{ruledtabular}
\end{table*}

\begin{table*}
\caption{\label{Table2} Fit results for Au-Au collisions at $\sqrt{s_{NN}}=200$ GeV and the measurement at central rapidity, $\mid y \mid < 0.35$. }
\begin{ruledtabular}
\begin{tabular}{ccccccc} \hline
\\
 \multicolumn{7}{c}{\textbf{with light nuclei}, $n_{dof}=18$}
\\
\hline\hline
   $\rho_1$ & $\rho_2$ & $T_{ch}$ (MeV) & $\mu_B$ (MeV) & $V$ (fm$^3$)& $\chi^{2}/n_{dof}$ & $\emph{p-value}\;(\%)$
\\
 \cline{1-7}
 0.181 & 0.380 & $157.9 \pm 0.6$ & $28.2 \pm 3.5$ & $1878.5 \pm 19.4$ & 1.64 & 4.2
\\
\hline\hline
 \multicolumn{7}{c}{parameter correlations}
\\
\hline
 \multicolumn{3}{c}{($T_{ch},\mu_B$)} & \multicolumn{2}{c}{($T_{ch},V$)} & \multicolumn{2}{c}{($\mu_B,V$)}
\\
\hline
 \multicolumn{3}{c}{-0.11} & \multicolumn{2}{c}{-0.84} & \multicolumn{2}{c}{0.59}
\\
\hline\hline
   $\rho_1$ & $\rho_2$ & $T_{ch}$ (MeV) & $\mu_B$ (MeV) & $V$ (fm$^3$)& $\chi^{2}/n_{dof}$ & $\emph{p-value}\;(\%)$
\\
\hline
 0.0 & 0.0 & $156.2 \pm 1.2$ &  $29.1 \pm 3.2$ & $2181.3 \pm 173.4$ & 1.70 & 3.2
\\
\hline\hline
 \multicolumn{7}{c}{parameter correlations}
\\
\hline
 \multicolumn{3}{c}{($T_{ch},\mu_B$)} & \multicolumn{2}{c}{($T_{ch},V$)} & \multicolumn{2}{c}{($\mu_B,V$)}
\\
\hline
 \multicolumn{3}{c}{0.12} & \multicolumn{2}{c}{-0.96} & \multicolumn{2}{c}{-0.11}
\\
\hline\hline
 \multicolumn{7}{c}{\textbf{without light nuclei}, $n_{dof}=10$}
\\
\hline
   $\rho_1$ & $\rho_2$ & $T_{ch}$ (MeV) & $\mu_B$ (MeV) & $V$ (fm$^3$)& $\chi^{2}/n_{dof}$ & $\emph{p-value}\;(\%)$
\\
\hline
 0.143 & 0.384 & $160.9 \pm 0.9$ & $24.0 \pm 8.8$ & $1759.1 \pm 45.4$ & 1.75 & 6.3
\\
\hline\hline
 \multicolumn{7}{c}{parameter correlations}
\\
\hline
 \multicolumn{3}{c}{($T_{ch},\mu_B$)} & \multicolumn{2}{c}{($T_{ch},V$)} & \multicolumn{2}{c}{($\mu_B,V$)}
\\
\hline
 \multicolumn{3}{c}{-0.407} & \multicolumn{2}{c}{-0.727} & \multicolumn{2}{c}{0.848}
\\
\hline\hline
   $\rho_1$ & $\rho_2$ & $T_{ch}$ (MeV) & $\mu_B$ (MeV) & $V$ (fm$^3$)& $\chi^{2}/n_{dof}$ & $\emph{p-value}\;(\%)$
\\
\hline
 0.0 & 0.0 & $163.4 \pm 2.3$ & $24.7 \pm 7.0$ & $1545.8 \pm 187.9$ & 1.50 & 13.1
\\
\hline\hline
 \multicolumn{7}{c}{parameter correlations}
\\
\hline
 \multicolumn{3}{c}{($T_{ch},\mu_B$)} & \multicolumn{2}{c}{($T_{ch},V$)} & \multicolumn{2}{c}{($\mu_B,V$)}
\\
\hline
 \multicolumn{3}{c}{0.041} & \multicolumn{2}{c}{-0.978} & \multicolumn{2}{c}{-0.033}
\\
\hline
\end{tabular}
\end{ruledtabular}
\end{table*}

\section{Conclusions}
\label{conclus}

 It is understood that from results of such a simple and speculative model no definite conclusions can be drawn/fomulated, so the following summary should be treated as the indication of what could happen if the very detailed correlation assumptions at least approximated the actual circumstances.

For collision examples considered here, non-zero correlations cause the slight increase of the the chemical freeze-out temperature and the decrease of the volume, only in the RHIC-STAR case without light nuclei the different behavior is observed, i.e. the temperature decreases and the volume increases. However one should notice that the $1\sigma$ intervals of temperatures determined for correlated and corresponding non-correlated cases overlap, so it is also likely that the temperature does not change. The opposite behavior of temperature and volume is confirmed by values of their correlation coefficients, all in the range from -0.98 to -0.73, what proves that they are highly anti-correlated. This is understandable because the number of particles is the increasing function of both temperature and volume, so if it is fixed at a given value (measured multiplicity), the increase of the volume enforces the decrease of the temperature and vice versa. The baryochemical potential stays practically unchanged, when correlations are switched on, only in the ALICE-LHC case with light nuclei, it changes by one order but still within errors. The correlations of baryochemical potential with other parameters are significant only for the STAR-RHIC case with $\rho_1, \rho_2 > 0$,  for ($T_{ch},\mu_B$) pair only in the case without light nuclei.

It should also be noticed that the inclusion of the experimental correlations in the presented toy model has determined the same freeze-out temperature, $\approx 158$ MeV, in both STAR-RHIC and ALICE-LHC cases. For the case without nuclei the temperature agrees within errors and is $\approx 160$ MeV. However, also for zero correlations but with inclusion of light nuclei into fits the same freeze-out temperature is obtained, $\approx 156$ MeV, so slightly lower. This is remarkable because, unexpectedly, the smaller  freeze-out temperature for LHC was found \cite{Stachel:2013zma} with respect to RHIC \cite{Andronic:2005yp}. That fact is confirmed also in this analysis: for the case without correlations and light nuclei, the freeze-out temperature for ALICE-LHC is 6 MeV lower than the corresponding temperature for STAR-RHIC.

Of course, these quantitative results are not definite, they only show the possibility.
To obtain more certain results, the analysis should be repeated with the values of correlation coefficients or elements of the covariance matrix somehow measured or estimated (e.g. with the help of Monte Carlo simulations) by experimentalists.

\begin{acknowledgments}
The author would like to thank Cezary Juszczak for help in editing the paper.
\end{acknowledgments}


\end{document}